\documentclass[conference]{IEEEtran}
\IEEEoverridecommandlockouts
\usepackage{cite}
\usepackage{amsmath,amssymb,amsfonts}
\usepackage{epsfig}
\usepackage{algorithmic}
\usepackage{graphicx}
\usepackage{textcomp}
\usepackage{xcolor}
\def\BibTeX{{\rm B\kern-.05em{\sc i\kern-.025em b}\kern-.08em
    T\kern-.1667em\lower.7ex\hbox{E}\kern-.125emX}}

\usepackage{multirow}
\usepackage{hhline}
\usepackage{tabularx}
\usepackage{float}
\usepackage{algorithm2e}

\usepackage{caption}
\usepackage{subcaption}
\usepackage{graphicx}
\usepackage{enumitem}
\usepackage{xcolor}
\usepackage{amsmath,amssymb}
\usepackage{hyperref}
\hypersetup{
    colorlinks=true,
    linkcolor=blue,
    filecolor=blue,      
    urlcolor=blue,
    pdftitle={Overleaf Example},
    pdfpagemode=FullScreen,
    }
\begin{document}

% Example definitions.
% --------------------
\def\x{{\mathbf x}}
\def\L{{\cal L}}
% \newcommand\ie{i.e.}
% \newcommand\eg{e.g.}
% \newcommand{\zhihao}[1]{{\color{orange} #1}}
% \newcommand{\todo}[1]{{\color{red} #1}}

% Title.
% ------
\title{Flexible Mixed Precision Quantization for Learned Image Compression}
%
% Single address.
% ---------------
% \name{Anonymous ICME submission}
% % \name{Md Adnan Faisal Hossain, Zhihao Duan, Fengqing Zhu}

% %Address and e-mail should NOT be added in the submission paper. They should be present only in the camera ready paper. 
% \address{}

\author{\IEEEauthorblockN{Md Adnan Faisal Hossain, Zhihao Duan, Fengqing Zhu}
\IEEEauthorblockA{\textit{Elmore Family School of Electrical and Computer Engineering} \\
\textit{Purdue University}\\
West Lafayette, IN, 47906, USA \\
\{hossai34, duan90, zhu0\}@purdue.edu}}

\maketitle

\begin{abstract}
Despite its improvements in coding performance compared to traditional codecs, Learned Image Compression (LIC) suffers from large computational costs for storage and deployment. Model quantization offers an effective solution to reduce the computational complexity of LIC models. However, most existing works perform fixed-precision quantization which suffers from sub-optimal utilization of resources due to the varying sensitivity to quantization of different layers of a neural network. In this paper, we propose a Flexible Mixed Precision Quantization (FMPQ) method that assigns different bit-widths to different layers of the quantized network using the fractional change in rate-distortion loss as the bit-assignment criterion. We also introduce an adaptive search algorithm which reduces the time-complexity of searching for the desired distribution of quantization bit-widths given a fixed model size. Evaluation of our method shows improved BD-Rate performance under similar model size constraints compared to other works on quantization of LIC models. We have made the source code available at \href{https://gitlab.com/viper-purdue/fmpq.git}{gitlab.com/viper-purdue/fmpq}.
\end{abstract}


\section{Introduction}
\label{sec:intro}
% \vspace{-0.2cm}

% \todo{
% Brief introduction to neural image compression as a form of lossy image compression. Recent performance of neural image compression surpassing many traditional codecs.

% Large model size of neural image compression models acts as an issue for large scale deployment and for deployment in resource constrained devices.

% Brief introduction to model compression techniques; model quantization and model pruning. Model compression techniques can act as a solution for reducing model size of neural image compression models while maintaining rate-distortion performance.

% We propose a multi-domain model compression technique combining mixed precision quantization with filter pruning.
% }

Recent developments of Learned Image Compression (LIC) methods \cite{balle2018variational, minnen2018joint, cheng2020learned, duan2023lossy, duan2023qarv, jiang2023mlic} have shown superior coding performance compared to traditional codecs such as JPEG, BPG, and VTM. However, these LIC models typically have $200$ to $400$ million parameters \cite{duan2023qarv}, leading to large memory requirements that make them difficult to deploy on resource-constrained devices. Existing works on the quantization of LIC models \cite{balle2018integer, hong2020efficient, sun2020end, sun2021learned, sun2022q, shi2023rate, jeon2023integer} attempt to tackle this problem by employing neural network quantization to convert the model parameters to a lower precision representation (often \textit{float32} to \textit{int8}). Furthermore, it has been shown in \cite{balle2018integer} that quantizing LIC models to integer precision also reduces decoding failures from cross-platform numerical round-off errors. 

\begin{figure}[t]
 \centerline{\epsfig{figure=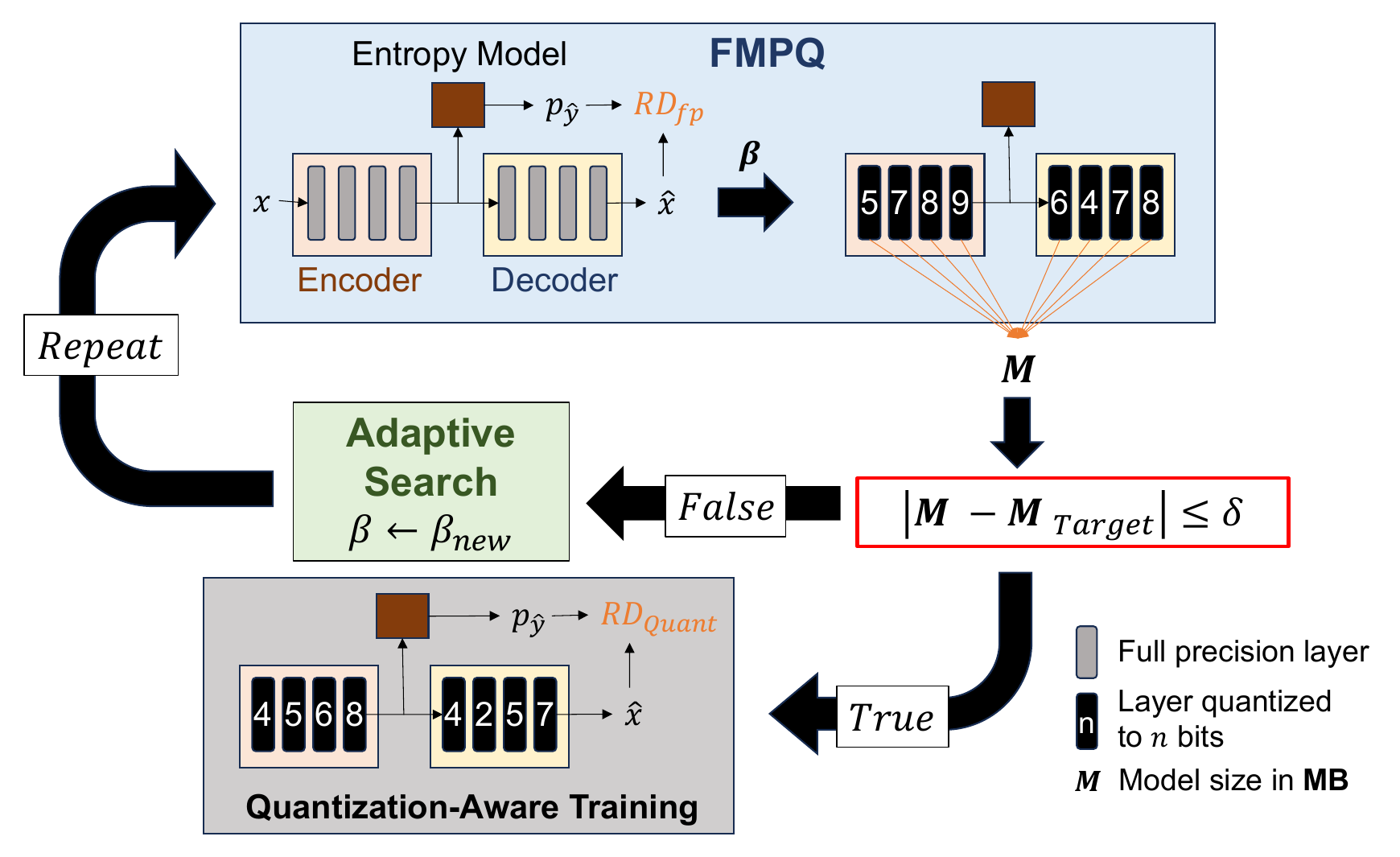, width=0.95\linewidth}}
% \caption{ConvNext Adaptive Layer Normalization block used to train a variable-rate feature compression model.}
% \vspace{-0.2cm}
\caption{Overview of our proposed Flexible Mixed-Precision Quantization (FMPQ) scheme. Using an initial value of the parameter $\beta$, mixed-precision quantization is performed. The quantized model size $M$ is computed, and if it is within an acceptable range $\delta$ from the desired model size $M_{\textit{Target}}$, quantization-aware training is performed. Otherwise, the adaptive search algorithm adjusts $\beta$, and the quantization process is repeated.}
\vspace{-0.2cm}
\label{fig:overview}
\end{figure}

These existing works assign a uniform bit-precision for each layer of the network. However, this can be sub-optimal as different layers of the network show different performance degradation to quantization. Mixed-precision quantization, a technique that assigns different bit-widths to different layers of the network, has been widely adopted in the existing literature on model quantization \cite{cai2020rethinking, sun2022entropy, liu2022instance, shi2023lossy} to solve this problem, but is yet to be explored for quantizing LIC models. In this work, we propose a Flexible Mixed Precision Quantization (FMPQ) technique coupled with an adaptive search algorithm to efficiently seek a distribution of bit widths that can maximize model performance given a fixed model size. An overview of our proposed method is shown in Fig.~\ref{fig:overview}. Our contributions include:
\setlist{nolistsep}
\begin{itemize}[noitemsep]
  \item We propose a Flexible Mixed-Precision Quantization (FMPQ) method for LIC models that utilizes the fractional change in rate-distortion loss (RD-Loss) as the bit-assignment criterion.
  \item We describe a quantization-aware training (QAT) scheme for LIC models that uses only the RD-Loss as the optimization function.
  \item We propose an adaptive search algorithm to reduce the time complexity of seeking the desired distribution of bit-widths across the neural network layers under the constraint of a fixed model size. 
\end{itemize}

\section{Preliminaries}
% \vspace{-0.1cm}

\subsection{Learned Image Compression}
\label{sec:preliminaries_1}
% \vspace{-0.2cm}
Learned image compression is a form of transform coding where the input image $x$ is transformed by an encoder $g_a$ to a compressed domain representation $y=g_a(x)$. $y$ is subsequently quantized to $\hat{y}$ and then compressed by lossless entropy coding using a prior distribution $p_{\hat{y}}(\hat{y})$. The quantized, compressed representation $\hat{y}$ is losslessly decoded at the decoder $g_s$ and used to reconstruct the image $\hat{x}=g_s(\hat{y})$. Many LIC models \cite{balle2018variational, minnen2018joint, cheng2020learned} also have a hyper encoder $h_a$ to encode side information $z=h_a(y)$ that needs to be compressed and transmitted to the decoder as well to facilitate entropy decoding. For such cases, the prior distribution is parametric, $p_{\hat{y}}(\hat{y}; \mu,\sigma)$ with parameters $(\mu,\sigma)=h_s(\hat{z})$. Here, $h_s$ is referred to as the hyper decoder. These LIC models are trained on the Rate-Distortion (RD) loss shown in Eq. \eqref{eqn:1}:
% \vspace{-0.2cm}
% \begin{equation}
% L_{RD} = Rate + \lambda \cdot Distortion \label{eqn:1}
% \end{equation}
% \begin{multline}  \label{eqn:2}
% L_{RD} = \mathbb{E}_{X \sim p_x} [-\log_{2}p_{\hat{y}|z}(\hat{y}|z) - \\ \log_{2}p_{\hat{z}}(\hat{z})] + \lambda \cdot d(x, \hat{x})  
% \end{multline}

\vspace{-0.2cm}
\begin{equation}  \label{eqn:1}
\begin{split}
L_{\text{RD}} & = Rate + \lambda \cdot Distortion \\
& = \mathbb{E}_{X \sim p_x} [-\log_{2}p_{\hat{y}|z}(\hat{y}|z) - \log_{2}p_{\hat{z}}(\hat{z})] + \lambda \cdot d(x, \hat{x})  
\end{split}
\end{equation}
where $\lambda$ is the Lagrangian multiplier used to obtain a trade-off between the rate and distortion term.

% \vspace{-0.2cm}
\subsection{Model Quantization}
\label{sec:preliminaries_2}
% \vspace{-0.2cm}
% Model quantization was first proposed in \cite{han2015deep}. Since then there has been a substantial body of work aimed at developing low precision integer networks. 
% Model quantization is a model compression technique that uses lower precision fixed-point representations as opposed to higher precision floating-point representations to store model parameters.  
The body of work on quantization can be grouped into Post-Training Quantization (PTQ) \cite{nagel2020up, wei2022qdrop, shi2023rate} and Quantization-Aware Training (QAT) \cite{esser2019learned, lee2021network, jeon2023integer}. While PTQ only requires a small number of calibration data and no retraining, QAT offers better model performance at the cost of retraining and through access to a larger dataset.  

Most approaches for quantization adopt uniform quantizers which scale the dynamic range of the weights and then quantize them to integers. The range of the scaled weights is determined by the quantization bit-width $b$. The lower the quantization bit-width, the higher the achieved compression at the cost of model performance. There is also some interesting work in non-uniform quantization  \cite{jeon2022mr}, but they are challenging to deploy on hardware and hence have limited practical applications. In our work, we adopt uniform quantization and formulate the quantization of weights or activations $x$ into quantized weights or activations $\hat{x}$ as described in Eq.~\eqref{eqn:2}:
\vspace{+0.2cm}
\begin{equation}  \label{eqn:2}
% \begin{split}
% \hat{w} & = s_{w} \times \left\{\left\lfloor clip\left( \frac{w}{s_{w}}+z_{w}, 0, 2^{b}\right) \right\rceil - z_{w} \right\}  \\
% \hat{a} & = s_{a} \times \left\{\left\lfloor clip\left( \frac{a}{s_{a}}+z_{a}, 0, 2^{b}\right)  \right\rceil - z_{a} \right\} 
\hat{x} = s \times \left\{\left\lfloor clip\left( \frac{x}{s}+z, 0, 2^{b}\right)  \right\rceil - z \right\} 
% \end{split}
\vspace{+0.2cm}
\end{equation}
where $\lfloor \cdot \rceil$ refers to the integer rounding operation. $x$ is first scaled by quantization parameters $s$ and $z$ and then clipped to the range $\left[0, 2^{b} \right]$. After that, it is rounded to integers before being scaled back to its original dynamic range. 

% The $s$ and $z$ for the weights are learnable parameters (static during inference) trained on the RD-loss, while the $s$ and $z$ for activations are determined dynamically during inference. 

% We further utilize channel-wise quantization as opposed to layer-wise quantization as it reduces quantization error at the cost of very small increase in storage requirements. The extra storage requirements for channel-wise quantization is due to the need for a set of quantization parameters for each filter of a network layer as opposed to laye-wise quantization which assigns only one set of quantization parameters to each layer.

% In our work we set both $s_{w}$ and $z_{w}$ as learnable parameters which are trained on the RD-loss. On the other hand $s_{a}$ and $z_{a}$ are dynamically determined during inference.
% \vspace{-0.2cm}
% \subsection{Mixed Precision Quantization}
% \label{sec:preliminaries_3}
% \vspace{-0.2cm}

\subsection{Mixed Precision Quantization}
\label{sec:preliminaries_3}
% \vspace{-0.2cm}

Quantizing all the layers of a neural network to the same precision suffers from sub-optimal resource utilization as it fails to account for the varying sensitivities of different layers to quantization. Mixed Precision Quantization addresses this problem by assigning different bit-widths to different layers of the neural network based on a bit assignment criterion.

% This gives rise to the problem of developing a criterion based on which the bit-width of each layer of the network can be determined such that a performance metric is maximized under the constraints of an overall bit-budget. 

Recent works like \cite{sun2022entropy} and \cite{shi2023lossy} estimate the entropy of the weights of each layer of the network and allocate the bit-width accordingly. Although reducing the entropy value of the weights can reduce the quantization error, it does not guarantee lower RD-loss, and hence can lead to a sub-optimal solution as shown in \cite{shi2023rate}. At the same time, the method from \cite{shi2023lossy} requires retraining of the model which can be impractical for off-the-shelf deployment. We propose to address these problems by employing the target metric we want to reduce (RD-loss in our case) as the criterion for determining the bit-width assigned to each layer. Our approach is a training-free method that leverages an efficient adaptive search algorithm to efficiently find the desired distribution of bit-widths. 

\begin{figure}[t]
 \centerline{\epsfig{figure=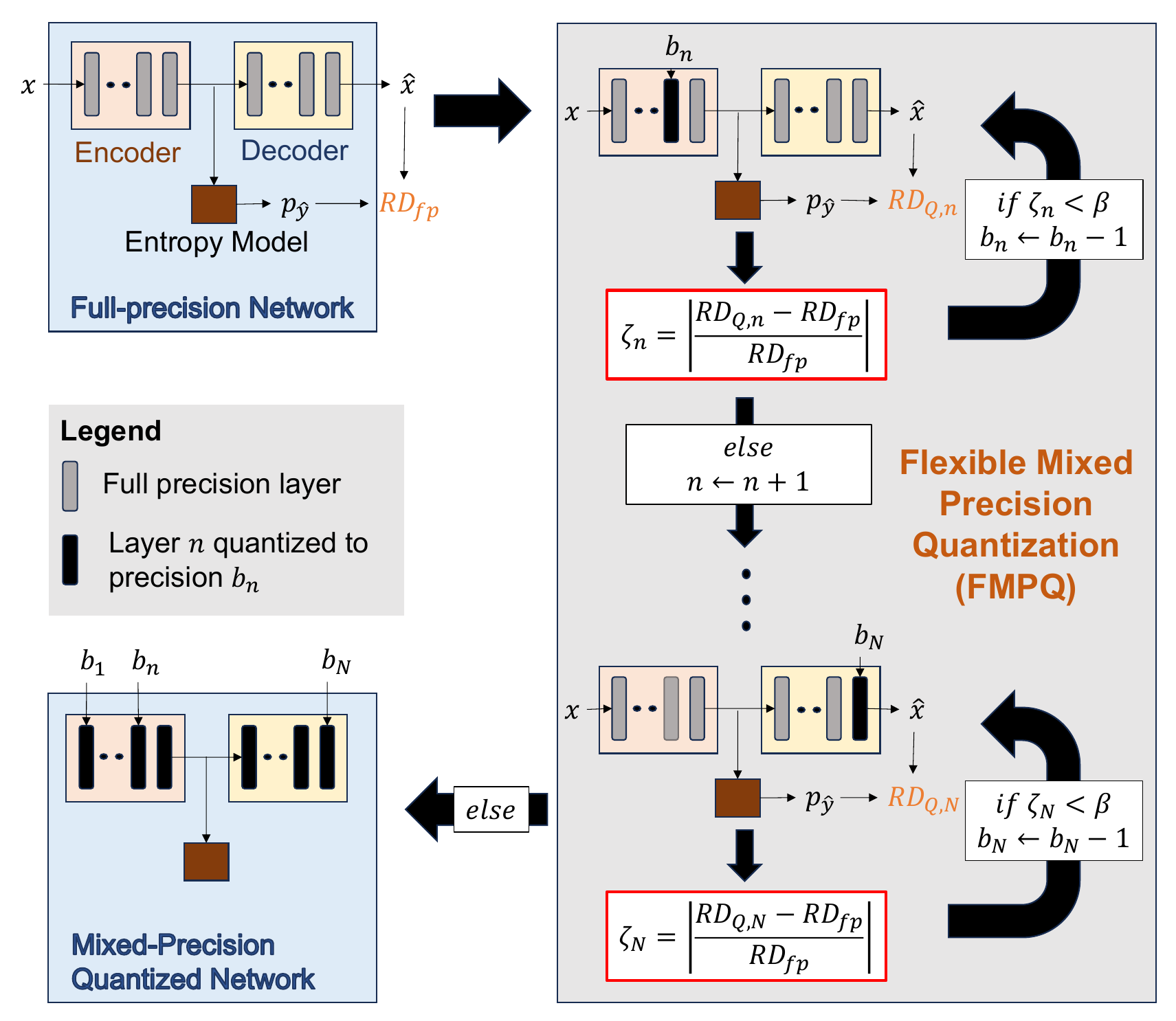, width=1.0\linewidth}}
% \caption{ConvNext Adaptive Layer Normalization block used to train a variable-rate feature compression model.}
% \vspace{-0.2cm}
\caption{Our proposed Flexible Mixed-Precision Quantization (FMPQ) method. To determine the bit-width $b_n$ assigned to the $n^{th}$ layer of the LIC model, $\zeta_n(b)$ is calculated using each candidate bit-width $b$ from $L$. The smallest bit-width from $L$ that satisfies $\zeta_n(b) < \beta$, is set as the value of $b_n$. This process is carried out seperately for all of the $N$ layers of the network.}
\label{fig:MPQ}
\vspace{-0.3cm}
\end{figure}

% \vspace{-0.3cm}
% \subsection{Model Pruning}
% \label{sec:preliminaries_4}
% \vspace{-0.2cm}

% Model pruning removes redundant weights from neural networks based on a predefined pruning criterion and pruning ratio. Pruning can be primarily divided into two categories based on the pruning granulairty: weight pruning \cite{han2015learning} and filter pruning \cite{molchanov2019importance}. Weight pruning focuses on selecting the weights to be pruned within a layer based on the individual importance of each weight element. On the contrary, filter pruning removes entire filters from a layer based on an average importance criterion measured using all the weight elements of a particular filter. In our work, we perform filter pruning as weight pruning leads to unstructured sparsity patterns which requires the use of complicated hardware support to achieve model compression. The pruning criterion is either weight-dependent such as filter norm \cite{he2018soft} or feature based like average rank of activation feature maps \cite{lin2020hrank}. 

% The criterion can have an intra-channel or inter-channel perspective, where the inter-channel perspective although is more costly to compute gives the pruning method access to cross-channel information that can further aid in identifying the redundant weights to be pruned.

% \vspace{-0.2cm}
\section{Method}
% \vspace{-0.1cm}

In Sec.~\ref{sec:method_1}, we describe our Flexible Mixed-Precision Quantization (FMPQ) method that employs the percentage change in RD-Loss as the bit-assignment criterion. We then explain the quantization-aware training scheme that we adopt for training our quantized network in Sec.~\ref{sec:method_2}. Finally, in Sec.~\ref{sec:method_3} we outline our proposed adaptive search algorithm that reduces the time-complexity of searching for the optimal distribution of bit-widths for a given model size.

% \vspace{-0.3cm}
\subsection{Flexible Mixed Precision Quantization Scheme}
\label{sec:method_1}
% \vspace{-0.2cm}

Our proposed FMPQ method shown in Fig.~\ref{fig:MPQ} assesses the sensitivity to quantization of each layer of the neural network. Subsequently, a specific bit-width from a set of candidate bit-widths $L = \{b : b \in \mathbb{N},~2 \leq b \leq b_{max} \}$, is assigned to each layer of the network based on the layer’s sensitivity to quantization. Then, each layer is quantized using its specific bit-width and finally quantization-aware training is performed.

% $L =\{2,3,4, ... b_{max} \}$

The fractional change in RD-loss, $\zeta_n(b)$ due to the quantization of the $n^{th}$ layer of the network by $b$ bits is used as the metric for determining the sensitivity to quantization of that layer. 
% \vspace{-0.2cm}
\begin{equation}  \label{eqn:3}
\zeta_n(b) =  \left| \dfrac{RD_{\text{quantized}, n}(b) - RD_{\text{full-precision}}}{RD_{\text{full-precision}}} \right|
\vspace{+0.3cm}
\end{equation}
$RD_{\text{full-precision}}$ refers to the rate-distortion loss of the full-precision floating point model on a calibration dataset of images $D_{\text{calib}}$. $RD_{\text{quantized}, n}(b)$ refers to the rate-distortion loss of an identical, full precision model with only the $n^{th}$ layer quantized using bit-width $b$.

We also define a threshold of tolerance $\beta$, on $\zeta_n$ to determine the lowest possible bit-width that can be assigned to the $n^{th}$ layer of the network under the constraints of a maximum tolerable RD-Loss. Our proposed algorithm calculates $\zeta_n$ using each bit-width from $L$ and assigns the lowest bit-width that satisfies the condition $\zeta_n < \beta$ as the bit-width for layer $n$. We can achieve a trade-off between RD-loss and bit-precision by varying $\beta$, where increasing $\beta$ increases the RD-loss but lowers the assigned bit-width. The proposed method for realizing the desired distribution of bit-widths $B_{\theta}$ for the LIC model is summarized in \textbf{Algorithm~\ref{alg:alg1}}. 

\RestyleAlgo{ruled}
\SetKw{KwInput}{Input:}
\SetKw{KwOutput}{Output:}
\SetKw{KwBuffer}{Buffer:}
\SetKw{Kwalgstart}{Initialize}
\SetKw{KwBy}{by}
\begin{algorithm}[hbt!]
	\caption{Mixed Precision Quantization}\label{alg:alg1}
        % \textbf{Input:} {Full precision $N$ layer floating point model $G_{\theta}$  with set of weights $\theta=\{\theta_1,\theta_2, ... \theta_N\}$, set of quantization parameters $\phi=\{\phi_1, \phi_2, ... \phi_N\}$, $D_{calib}$, $\beta$, $L$} \\
        % \textbf{Output:} {set of assigned bit-precision for each quantized layer  $B_{\theta}={b_{\theta_1},b_{\theta_2}, ... b_{\theta_N})  }$ where each $b_{\theta_i} \in L$}\\
    \KwInput{Full precision $N$ layer floating point model $G_{\theta}$  with set of weights $\theta=\{\theta_1,\theta_2, ... \theta_N\}$, set of quantization parameters $\phi=\{\phi_1, \phi_2, ... \phi_N\}$, $X \sim D_{\text{calib}}$, $\beta$, $L$} \\
    \KwBuffer{Quantized weight of $n^{th}$ layer $\hat{\theta}_n$} \\
    \KwOutput{$B_{\theta}=[b_{\theta_1},b_{\theta_2}, ... b_{\theta_N}]$ where each $b_{\theta_i} \in L$}\\
    \Kwalgstart $B_{\theta} \gets [b_{\text{max}},b_{\text{max}}, ... b_{\text{max}}]$\;
    $RD_{\text{fp}} = L_{RD}(X, \theta)$\;
    \For{$n=0$ \KwTo $N$ \KwBy $1$}{
    \For{$b=b_{\text{max}}$ \KwTo $2$ \KwBy $-1$}{
    $b_{\theta_n} \gets b$\;
    Calibrate $\phi$ w.r.t $D_{\text{calib}}$ using $B_{\theta}$\\
    $\hat{\theta}_n \gets \text{Quantize}(\theta_n; b_{\theta_n})$\;
    $\hat{\theta} = \{\theta_1, \theta_2, ... \hat{\theta}_n, ...  \theta_N\}$\;
    $RD_{\text{Q, n}} = L_{RD}(X, \hat{\theta})$\;
    $\zeta_n =  \left| \frac{RD_{\text{Q, n}} - RD_{\text{fp}}}{RD_{\text{fp}}} \right| \times 100$\;
    \If{$\zeta_n \geq \beta$}{
    $b_{\theta_n} \gets b - 1$\;
    \textbf{break}
    }
    }
    }
\end{algorithm} 
\vspace{-0.2cm}

\subsection{Quantization Aware Training on the RD-Loss}
\label{sec:method_2}
% \vspace{-0.2cm}

Once the bit-width for each layer of the network has been determined, we perform quantization-aware training. Following the method in \cite{esser2019learned}, we train both the neural network weights and quantization parameters. We choose the RD-loss as the loss function for quantization-aware training as it directly optimizes over the quantities of interest in image compression (bit-rate and image distortion). We also adopt the leaky-clip module of \cite{jeon2023integer} during training to address the problem of vanishing gradients due to the clipping function. However, unlike \cite{jeon2023integer} we do not train our model on the quantization error loss since it has been shown in \cite{shi2023rate} that a larger quantization error may sometimes lead to lower RD-loss.

We set the quantization parameters for the weights $s_w$ and $z_w$ as learnable parameters (static during inference), trained on the RD-loss, while the quantization parameters for the activations $s_a$ and $z_a$ are determined dynamically during inference. We further utilize channel-wise quantization as opposed to layer-wise quantization as it reduces quantization error at the cost of minimal increase in storage requirements. The extra storage requirements for channel-wise quantization is due to the need for a set of quantization parameters for each filter in a layer as opposed to layer-wise quantization which assigns only one set of quantization parameters to the whole layer.
% \vspace{-0.3cm}

\subsection{Adaptive Search Algorithm}
\label{sec:method_3}
% \vspace{-0.2cm}

The FMPQ method outlined in Sec.~\ref{sec:method_1} can achieve a trade-off between RD-loss and model size by varying the value of $\beta$. However, we still need a method to find the value of $\beta$ that achieves our desired model size. A naive solution is to perform an exhaustive search with a fixed step-size over $\beta$ to obtain a quantized model that satisfies our model size constraints, but the cost of such a search grows exponentially as shown in the third column of Table~\ref{tab:tab1}. Therefore, we propose an adaptive search algorithm to determine the desired $B_{\theta}$ that satisfies a given model size constraint. 

We define our desired model size in terms of a target compression ratio $CR_{\text{target}}$, which is the ratio of the desired model size in \textbf{MB} of the mixed-precision quantized model to the model size in \textbf{MB} of an 8-bit fixed precision quantized model with same network architecture. Our search algorithm is a variable-step size search that adaptively changes the increment applied to $\beta$ at each search step based on the absolute difference between $CR_{\text{target}}$ and the achieved compression ratio at that step $CR$. Therefore, when $|CR - CR_{\text{target}}|$ is relatively large, larger increments are applied to $\beta$, and when $|CR - CR_{\text{target}}|$ is relatively small, smaller increments are applied. If at any step $CR$ is below the target compression ratio $CR_{\text{target}}$, the increment applied in the past step is reversed and the algorithm proceeds with a smaller increment size. The search is terminated when the achieved compression ratio $CR$ is within a certain small threshold ($0.01$) of the target compression ratio $CR_{\text{target}}$. The factors by which search increments are modified are determined empirically such that the algorithm converges efficiently for different target compression ratios as shown in the second column of Table~\ref{tab:tab1}. A summary of our proposed adaptive search algorithm is described in \textbf{Algorithm~\ref{alg:alg2}}.

% If at any step $CR$ is below the target compression $CR_{target}$, the increment applied in the past step is reversed and the algorithm proceeds with a smaller increment size.

% \vspace{-0.2cm}
% \subsection{Pruning based on activation feature maps}
% \label{sec:method_3}
% \vspace{-0.2cm}

% Recent works have shown that pruning neural network filters based on the importance of corresponding feature maps yields better results than those that are based on the importance of the filter weights \cite{lin2020hrank, tang2020scop, sui2021chip}. Therefore, we choose to model after the feature guided filter pruning method.

\begin{algorithm}[hbt!]
	\caption{Adaptive Search for $\beta$}\label{alg:alg2}
    \KwInput{Initial value of threshold of tolerance on RD-loss $\beta_{\text{init}}$, $CR_{\text{target}}$,} \\
    \KwBuffer{Threshold of tolerance for current iteration, $\beta$, increment applied to $\beta$ at current iteration, $\alpha_{\text{beta}}$, compression ratio $CR$}\\
    \KwOutput{$B_{\theta}$ that achieves $CR_{\text{target}}$}\\
    \Kwalgstart $\alpha_{\beta} \gets 1, \beta \gets \beta_{\text{init}}$\;
    \While{True}{
    Execute \textbf{Algorithm \ref{alg:alg1}} to get $B_{\theta}$ for current step\;
    $CR =  \dfrac{\textbf{model}(\hat{\theta}, B_{\theta})}{\textbf{model}(\hat{\theta}, \{8, 8, ... ... 8\})} $\;
    \uIf{$\left|CR - CR_{\text{target}} \right| \leq 0.01$}{
    \textbf{break}\;
      }
      \uElseIf{$CR \leq CR_{\text{target}}$}{
        $\beta \gets \text{maximum}(\beta - \alpha_{\beta}, 10^{-3})$\;
        $\alpha_{\beta} \gets \alpha_{\beta} \times 0.1$\;
        $\beta \gets \beta + \alpha_{\beta}$\;
      }
      \Else{
        \uIf{$\left|CR - CR_{\text{target}} \right| \geq 0.25$}{
        $\alpha_{\beta} \gets \alpha_{\beta} \times 5$\;
        }
        \uElseIf{$\left|CR - CR_{\text{target}} \right| \geq 0.10$}{
        $\alpha_{\beta} \gets \alpha_{\beta} \times 2$\;
        }
        $\beta \gets \beta + \alpha_{\beta}$\;
      }
    }
        \end{algorithm}

\section{Experiments}
\label{sec:exp}

We compare the performance of our Flexible Mixed Precision Quantization (FMPQ) method with 8-bit fixed precision quantization (FPQ) on three different LIC models \cite{balle2018variational, minnen2018joint, cheng2020learned}. Next, we showcase the flexible nature of our FMPQ method, and the time-complexity reduction achieved by our adaptive search algorithm. We then make a comparison of our proposed method with existing literature \cite{hong2020efficient, sun2021learned, shi2023rate, jeon2023integer, sun2022q}, and conclude by analyzing the distribution of bit-widths across our propopsed mixed-precision quantized model.

\subsection{Experimental Setup}
\label{sec:exp_1}
% \vspace{-0.2cm}

In all our experiments, the full-precision (float32) model is first quantized, and then fine-tuned using the quantization-aware training strategy of Sec.~\ref{sec:method_2}. We obtain the pre-trained weights of the \textit{Scale Hyperprior} \cite{balle2018variational} and \textit{Cheng Anchor 2020} \cite{cheng2020learned} LIC models from the publicly available CompressAI PyTorch Library \cite{begaint2020compressai}. We train a variant of the \textit{Mean Scale Hyperprior} \cite{minnen2018joint} LIC model that is used in \cite{jeon2023integer} to obtain the full precision baseline. The baseline model is trained using the COCO dataset \cite{lin2014microsoft} for 90 epochs with the Adam optimizer and a batch size of 16. We also apply a cosine learning rate decay with initial learning rate set to $10^{-4}$. We train six different baseline networks to obtain six different quality levels of compression corresponding to $\lambda=\{0.0018, 0.0035, 0.0067, 0.0130, 0.0250, 0.0483\}$.   

The relative Rate-Distortion performance of the quantized models in the following experiments are measured using the Bjontegaard delta rate (BD-Rate) with respect to the baseline full-precision models. We measure the compression achieved by the quantized models by comparing their model size in \textbf{MB} to that of the full-precision models. We use three different datasets, Kodak \cite{kodak1993kodak}, Tecnick \cite{asuni2013testimages} and Clic \cite{CLIC2020} to evaluate the BD-Rates. 

% \vspace{-0.1cm}
\subsection{Coding Performance of FMPQ vs 8-bit FPQ}
\label{sec:exp_2}
% \vspace{-0.2cm}

In this section, we demonstrate the performance of our Flexible Mixed Precision Quantization (FMPQ) method vs 8-bit Fixed Precision Quantization (FPQ) on three LIC models: \textit{Scale Hyperprior} \cite{balle2018variational}, \textit{Mean Scale Hyperprior} \cite{minnen2018joint} and \textit{Cheng Anchor 2020} \cite{cheng2020learned}. We obtain both the fixed-precision quantized and mixed-precision quantized models from the pre-trained floating point baselines. We only use $16$ images from the COCO dataset to create the calibration dataset, $D_{\text{calib}}$ for our FMPQ method. After quantization, we fine-tune both the quantized models using the COCO dataset for 30 epochs with the Adam optimizer and use a learning rate of $10^{-5}$ for the model parameters and learning rate of $10^{-4}$ for the quantization parameters.

For a fair comparison, we set the hyperparameter $CR_{\text{target}}$ defined in Sec.~\ref{sec:method_3} to $1$ so that the size of the model quantized using our proposed FMPQ method is similar to the size of the model undergoing 8-bit FPQ. We can see from Table~\ref{tab:tab2} that FMPQ can achieve a BD-Rate reduction of $0.96\% ([7.44-6.48]\%)$, $2.34\% ([3.54-1.20]\%)$ and $1.16\% ([2.05-0.89]\%)$ compared to 8-bit FPQ on the Kodak dataset using the \textit{Scale Hyperprior}, \textit{Mean Scale Hyperprior} and \textit{Cheng Anchor 2020} LIC models from \cite{balle2018variational}, \cite{minnen2018joint} and \cite{cheng2020learned} respectively. Similar results can be observed on the Tecnick and Clic datasets. Therefore, our proposed FMPQ method shows better Rate-Distortion performance compared to 8-bit FPQ while achieving similar model size reduction. The corresponding RD-curves can be found in Appendix~\ref{sec:appendix_1}
% \vspace{-0.2cm}

\begin{table}[t]
\begin{center}
\caption{Comparison of our adaptive search algorithm vs exhaustive search over $\beta$ (initial value of $0.01$ and fixed increments of $0.01$) using the \textit{Mean Scale Hyperprior Model} \cite{minnen2018joint}.} \label{tab:tab1}
% \vspace{-0.2cm}
\scalebox{1.15}{
\begin{tabular}{c|cc|c}
\hline
\multirow{2}{*}{\begin{tabular}[c]{@{}c@{}}Compression\\ Ratio\\ (CR)\end{tabular}} & \multicolumn{2}{c|}{\begin{tabular}[c]{@{}c@{}}No. of iterations \\ to converge\end{tabular}} & \multirow{2}{*}{\begin{tabular}[c]{@{}c@{}}Convergence\\ time difference\\ (minutes)\end{tabular}} \\ \cline{2-3}
 & \multicolumn{1}{c|}{\begin{tabular}[c]{@{}c@{}}Adaptive \\ Search\end{tabular}} & \begin{tabular}[c]{@{}c@{}}Exhaustive\\ Search \end{tabular} &  \\ \hline
0.99 & \multicolumn{1}{c|}{6} & 7 & 2.4 \\ 
0.90 & \multicolumn{1}{c|}{6} & 15 & 22.5 \\ 
0.85 & \multicolumn{1}{c|}{12} & 27 & 38.75 \\ 
0.75 & \multicolumn{1}{c|}{10} & 80 & 189 \\ 
0.65 & \multicolumn{1}{c|}{6} & 160 & 488 \\ 
0.60 & \multicolumn{1}{c|}{2} & 500 & 1,~245 \\ 
0.55 & \multicolumn{1}{c|}{6} & 650 & 2,~361 \\ 
0.50 & \multicolumn{1}{c|}{7} & 900 & 3,~317 \\ 
\hline
\end{tabular}}
\vspace{-0.1cm}
\end{center}
\end{table}

\begin{table}[t]
\begin{center}
\caption{BD-Rate vs Model Size (\textbf{MB}) comparison of proposed FMPQ method vs 8-bit FPQ.} \label{tab:tab2}
% \vspace{-0.2cm}
\scalebox{1.05}{
\begin{tabular}{c|c|c|c|c|c}
\hline
& \multirow{3}{*}{\begin{tabular}[c]{@{}c@{}}Quant.\\ Method\end{tabular}} & \multicolumn{3}{c|}{BD-Rate(\%)} & Model \\
\hhline{~~---~}
Model &  & Kodak & Tecnick & Clic & Size \\
\hhline{~~~~~~}
 &  & \cite{kodak1993kodak} & \cite{asuni2013testimages} & \cite{CLIC2020} & (MB) \\
\hline
\multirow{3}{*}{\begin{tabular}[c]{@{}c@{}}Scale\\ Hyperprior\\ \cite{balle2018variational}\end{tabular}} & None & 0 & 0 & 0 & 30.44 \\
& 8-bit FPQ & +7.44 & +9.11 & +8.95 & 7.66 \\
& FMPQ & +6.48 & +8.22 & +8.52 & 7.65 \\
\hline
\multirow{3}{*}{\begin{tabular}[c]{@{}c@{}}Mean Scale\\ Hyperprior\\ \cite{minnen2018joint}\end{tabular}} & None & 0 & 0 & 0 & 34.68 \\
& 8-bit FPQ & +3.54 & +5.87 & +5.78 & 8.71 \\
& FMPQ & +1.20 & +2.64 & +2.58 & 8.73 \\
\hline
\multirow{3}{*}{\begin{tabular}[c]{@{}c@{}}Cheng\\ Anchor 2020\\ \cite{cheng2020learned}\end{tabular}} & None & 0 & 0 & 0 & 76.98 \\
& 8-bit FPQ & +2.05 & +4.97 & +3.54 & 19.36 \\
& FMPQ & +0.89 & +2.68 & +1.70 & 19.26 \\
% \hline
% \multirow{3}{*}{\begin{tabular}[c]{@{}c@{}}Lu\\ 2022\\ \cite{lu2022high}\end{tabular}} & None & 0 & 0 & 0 & \todo{To do} \\
% & 8-bit FPQ & \todo{To do} & \todo{To do} & \todo{To do} & \todo{To do} \\
% & FMPQ & \todo{To do} & \todo{To do} & \todo{To do} & \todo{To do} \\
\hline
\end{tabular}}
% \vspace{-0.1cm}
\end{center}
\end{table}

\begin{table}[t]
\begin{center}
\caption{Trade-off between BD-Rate and Compression Ratio achieved by our FMPQ method.} \label{tab:tab3}
% \vspace{-0.2cm}
\scalebox{1.05}{
\begin{tabular}{c|c|c|c|c}
\hline
Model & $CR_{\text{target}}$ & \begin{tabular}[c]{@{}c@{}}BD-Rate\\ Kodak (\%)\end{tabular} & \begin{tabular}[c]{@{}c@{}}Model Size\\ (MB)\end{tabular} & \begin{tabular}[c]{@{}c@{}}Compression\\ Raio\end{tabular} \\
\hline
\multirow{4}{*}{\begin{tabular}[c]{@{}c@{}}Cheng\\ Anchor\\ 2020\end{tabular}} 
& 1.0 & +0.89  & 19.26 & 4 $\times$ \\
& 0.94 & +0.99  & 18.34 & 4.2 $\times$ \\
& 0.75 & +2.04  & 14.46 & 5.32 $\times$ \\
& 0.60 & +3.38  & 12.27 & 6.27 $\times$ \\
\hline

\end{tabular}}
\vspace{-0.4cm}
\end{center}
\end{table}

\subsection{Trade-off between BD-Rate and Model Size}
\label{sec:exp_3}
% \vspace{-0.2cm}

Our proposed FMPQ method can achieve a trade-off between model size in \textbf{MB} and BD-Rate by tuning the hyper-parameter $CR_{\text{target}}$ defined in Sec.~\ref{sec:method_3}. This is shown in Table~\ref{tab:tab3} (corresponding RD-curves in Appendix~\ref{sec:appendix_2}) using the \textit{Cheng Anchor 2020} \cite{cheng2020learned} model. By setting $CR_{\text{target}}$ to $1.0$, our mixed-precision quantized model has the same size as an 8-bit fixed-precision quantized model. In other words, it achieves $4 \times$ model size compression from the full-precision model. The model size compression can be subsequently increased at the cost of BD-Rate drop by reducing $CR_{\text{target}}$ as shown in Table~\ref{tab:tab3}. For example, decreasing $CR_{\text{target}}$ from $1.0$ to $0.60$ decreases model size by $6.99 (19.26-12.27)$ \textbf{MB}, at the cost of $2.49 \% ([3.38-0.89]\%)$ BD-Rate drop.

Given a fixed value of $CR_{\text{target}}$, our proposed adaptive search algorithm can significantly reduce the time required to search for the desired bit distribution. This is demonstrated in Table~\ref{tab:tab1}, where we can see that for a $CR_{\text{target}}$ of $0.75$, our adaptive search requires $8 \times$ less search steps and is $3$ hours faster than an exhaustive search. With an initial value of $\beta$ set to $1$, the time-complexity of exhaustive search increases exponentially as $CR_{\text{target}}$ is reduced, whereas the complexity of our adaptive search remains roughly constant. For example, when $CR_{\text{target}}$ is dropped from $0.99$ to $0.50$, our adaptive search only requires 1 more step to converge, but an exhaustive search would take $893$ more steps.

\begin{table}[t]
\begin{center}
\caption{Comparison of our proposed methods with other quantized LIC baselines. Results obtained using our proposed method are highlighted in bold.} \label{tab:tab6}
% \vspace{-0.2cm}
\scalebox{1.05}{
\begin{tabular}{c|c|c|c}
\hline
Model & Method & \begin{tabular}[c]{@{}c@{}}BD-Rate\\ Kodak (\%)\end{tabular} & \begin{tabular}[c]{@{}c@{}}Model Size\\ (MB)\end{tabular} \\
\hline
\multirow{2}{*}{\begin{tabular}[c]{@{}c@{}}Scale\\ Hyperprior\\ (used in \cite{hong2020efficient})\end{tabular}} & \begin{tabular}[c]{@{}c@{}}\textbf{FMPQ}\\ \textbf{(w=MP, a=10)}\end{tabular} & \textbf{+1.01} & \textbf{7.65} \\
& \begin{tabular}[c]{@{}c@{}}Method from \cite{hong2020efficient}\\(w=8, a=10)\end{tabular} & +26.5 & 7.64 \\
\hline
\multirow{2}{*}{\begin{tabular}[c]{@{}c@{}}Cheng\\ Anchor 2020\\ (used in \cite{shi2023rate})\end{tabular}} & \begin{tabular}[c]{@{}c@{}}\textbf{FMPQ}\\ \textbf{(w=MP, a=8)}\end{tabular} & \textbf{+0.89} & \textbf{19.26} \\
& \begin{tabular}[c]{@{}c@{}}Method from \cite{shi2023rate}\\(w=8, a=8)\end{tabular} & +4.88 & 19.36 \\
\hline
% \multirow{2}{*}{\begin{tabular}[c]{@{}c@{}}Lu\\ 2022\\ (used in \cite{shi2023rate})\end{tabular}} & \begin{tabular}[c]{@{}c@{}}\textbf{FMPQ}\\ \textbf{(w=MP, a=8)}\end{tabular} & \todo{To do} & \todo{To do} \\
% & \begin{tabular}[c]{@{}c@{}}Method from \cite{shi2023rate}\\(w=8, a=8)\end{tabular} & +4.88 & \todo{To do} \\
% \hline
% \multirow{2}{*}{\begin{tabular}[c]{@{}c@{}}Lu\\ 2022\\ (used in \cite{shi2023rate})\end{tabular}} & \begin{tabular}[c]{@{}c@{}}\textbf{FMPQ}\\ \textbf{(w=MP, a=8)}\end{tabular} & \todo{To do} & \todo{To do} \\
% & \begin{tabular}[c]{@{}c@{}}Method from \cite{shi2023rate}\\(w=8, a=8)\end{tabular} & +3.70 & \todo{To do} \\
% \hline
\multirow{5}{*}{\begin{tabular}[c]{@{}c@{}}\\ Mean\\ Scale\\ Hyperprior\\ (used in \\ \cite{jeon2023integer, sun2021learned, sun2022q})\end{tabular}} & \begin{tabular}[c]{@{}c@{}} \textbf{FMPQ}\\ \textbf{(w=MP, a=8)} \end{tabular} & \textbf{+1.20} & \textbf{8.73} \\
& \begin{tabular}[c]{@{}c@{}} Method from \cite{jeon2023integer}\\(w=8, a=8) \end{tabular} & -0.54 & 6.62 \\
& \begin{tabular}[c]{@{}c@{}}Method from \cite{sun2021learned}\\(w=8, a=8)\end{tabular} & +5.59 & 8.71 \\
% & \begin{tabular}[c]{@{}c@{}}\textbf{FMPQ}\\ \textbf{(w=MP, a=4)}\end{tabular} & \textbf{+46.65} & \textbf{3.49} \\
% & \begin{tabular}[c]{@{}c@{}}Method from \cite{jeon2023integer}\\(w=4, a=4)\end{tabular} & +56.31 & 3.49 \\
% \hhline{~---}
% & \begin{tabular}[c]{@{}c@{}}\textbf{FMPQ + pruning} \\ \textbf{(w=MP, a=8)}\end{tabular} & \textbf{+3.63} & \textbf{5.02} \\
& \begin{tabular}[c]{@{}c@{}}Method from \cite{sun2022q}\\(w=8, a=8)\end{tabular} & +4.98 & 8.71 \\
\hline

\end{tabular}}
\vspace{-0.3cm}
\end{center}
\end{table}

\subsection{Comparison with other Quantized LIC}
\label{sec:exp_5}
% \vspace{-0.2cm}

In this section, we compare our proposed \textbf{FMPQ} method with other works on quantizing LIC models. We make the comparisons using the same floating-point baseline LIC models as the original papers. The comparisons are shown in Table~\ref{tab:tab6}. We observe that compared with the method from \cite{hong2020efficient}, our method can achieve a $25.49\%$ BD-Rate gain while maintaining a similar model size as theirs. Our proposed method also outperforms the quantization method from \cite{shi2023rate}, achieving a BD-Rate gain of $3.99\%$ using the \textit{Cheng Anchor 2020} model as the full-precision baseline. Although our method can achieve a $4.39\% ([5.59-1.20]\%)$ BD-Rate gain compared to \cite{sun2021learned} and a $3.78\% ([4.98-1.20]\%)$ D-Rate gain compared to \cite{sun2022q} using the \textit{Mean Scale Hyperprior} model, it performs worse ($1.74\% ([1.20+0.54]\%)$ BD-Rate drop) than the QAT method from \cite{jeon2023integer}. However, it should be noted that our results are based on the quantization of six different networks corresponding to six quality levels of image compression, whereas they use only four.

% Finally, we compare our \textbf{FMPQ} method with the method from \cite{sun2022q} which is the only other work that performs joint quantization and pruning of LIC models. Our method also outperforms \cite{sun2022q}, achieving a BD-Rate gain of $1.35\%$ using a model that is $1.95$ \textbf{MB} smaller.

\subsection{Distribution of bit-precisions using FMPQ}
\label{sec:exp_6}
% \vspace{-0.2cm}

\begin{figure}[t]
 \centerline{\epsfig{figure=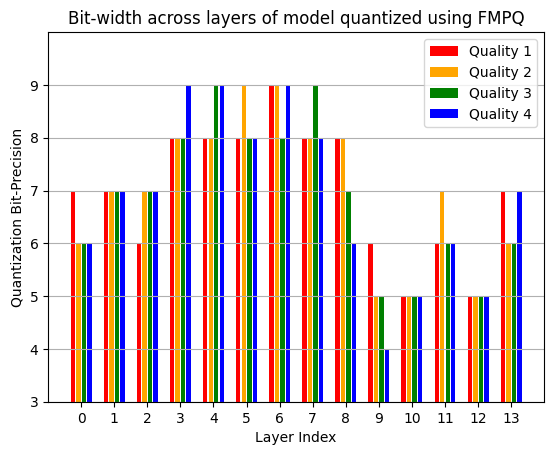, width=0.95\linewidth}}
% \caption{ConvNext Adaptive Layer Normalization block used to train a variable-rate feature compression model.}
\vspace{-0.1cm}
\caption{Distribution of quantization bit-widths using our FMPQ method on the \textit{Mean Scale Hyperprior} model.}
\vspace{-0.2cm}
\label{fig:ablation}
\end{figure}

We conduct a study of the distribution of bit-widths across the \textit{Mean Scale Hyperprior} model quantized using our FMPQ method. In Fig.~\ref{fig:ablation}, we plot the distribution for the four quantized networks corresponding to the four quality levels $\lambda=\{0.0018, 0.0035, 0.0067, 0.0130\}$. The \textit{Mean Scale Hyperprior} model has 14 quantized convolution and transposed convolution layers. Layers  $[0, 3]$ correspond to the main encoder, $[4, 7]$ correspond to the main decoder, $[8, 10]$ correspond to the hyper encoder and $[11, 13]$ correspond to the hyper decoder. We can observe from the figure that the weights in the main-path (layer $[0, 7]$) are in general more sensitive to quantization and require higher bit-widths as compared to the weights in the hyper-path. The last layer of the encoder (layer $3$) also requires a high bit-width as it contains most of the information about the latent representation of the image. The last layer of the decoder (layer $7$) generally also requires a high bit-width as it contains most of the information about the reconstructed output image.

\section{Conclusion}
\label{sec:conclusion}
% \vspace{-0.2cm}

In this paper, we propose a Flexible Mixed Precision Quantization (FMPQ) method for LIC models that utilizes the fractional change in rate-distortion loss as the bit-assignment criterion. Our method can be combined with an adaptive search algorithm to achieve a trade-off between BD-Rate performance and model size compression. We then demonstrate the superiority of our adaptive search algorithm compared to an exhaustive search in reducing the time complexity of finding the desired distribution of bit-widths for a given model size. We perform extensive experiments to show that our proposed FMPQ method can achieve better BD-Rate performance (with respect to the full-precision model) than 8-bit fixed precision quantization over three widely used image datasets. We finally compare our FMPQ method with other existing work on the quantization of LIC models and show that we are able to achieve better or similar performance.

% \vspace{-0.1cm}

\bibliographystyle{IEEEbib}
{\bibliography{icme2024main}}

\begin{thebibliography}{10}

\bibitem{balle2018variational}
Johannes Ball{\'e}, David Minnen, Saurabh Singh, Sung~Jin Hwang, and Nick Johnston,
\newblock ``Variational image compression with a scale hyperprior,''
\newblock {\em arXiv preprint arXiv:1802.01436}, 2018.

\bibitem{minnen2018joint}
David Minnen, Johannes Ball{\'e}, and George~D Toderici,
\newblock ``Joint autoregressive and hierarchical priors for learned image compression,''
\newblock {\em Advances in neural information processing systems}, vol. 31, 2018.

\bibitem{cheng2020learned}
Zhengxue Cheng, Heming Sun, Masaru Takeuchi, and Jiro Katto,
\newblock ``Learned image compression with discretized gaussian mixture likelihoods and attention modules,''
\newblock in {\em Proceedings of the IEEE/CVF conference on computer vision and pattern recognition}, 2020, pp. 7939--7948.

\bibitem{duan2023lossy}
Zhihao Duan, Ming Lu, Zhan Ma, and Fengqing Zhu,
\newblock ``Lossy image compression with quantized hierarchical vaes,''
\newblock in {\em Proceedings of the IEEE/CVF Winter Conference on Applications of Computer Vision}, 2023, pp. 198--207.

\bibitem{duan2023qarv}
Zhihao Duan, Ming Lu, Jack Ma, Yuning Huang, Zhan Ma, and Fengqing Zhu,
\newblock ``Qarv: Quantization-aware resnet vae for lossy image compression,''
\newblock {\em IEEE Transactions on Pattern Analysis and Machine Intelligence}, 2023.

\bibitem{jiang2023mlic}
Wei Jiang, Jiayu Yang, Yongqi Zhai, Peirong Ning, Feng Gao, and Ronggang Wang,
\newblock ``Mlic: Multi-reference entropy model for learned image compression,''
\newblock in {\em Proceedings of the 31st ACM International Conference on Multimedia}, 2023, pp. 7618--7627.

\bibitem{balle2018integer}
Johannes Ball{\'e}, Nick Johnston, and David Minnen,
\newblock ``Integer networks for data compression with latent-variable models,''
\newblock in {\em International Conference on Learning Representations}, 2018.

\bibitem{hong2020efficient}
Weixin Hong, Tong Chen, Ming Lu, Shiliang Pu, and Zhan Ma,
\newblock ``Efficient neural image decoding via fixed-point inference,''
\newblock {\em IEEE Transactions on Circuits and Systems for Video Technology}, vol. 31, no. 9, pp. 3618--3630, 2020.

\bibitem{sun2020end}
Heming Sun, Zhengxue Cheng, Masaru Takeuchi, and Jiro Katto,
\newblock ``End-to-end learned image compression with fixed point weight quantization,''
\newblock in {\em 2020 IEEE International Conference on Image Processing (ICIP)}. IEEE, 2020, pp. 3359--3363.

\bibitem{sun2021learned}
Heming Sun, Lu~Yu, and Jiro Katto,
\newblock ``Learned image compression with fixed-point arithmetic,''
\newblock in {\em 2021 Picture Coding Symposium (PCS)}. IEEE, 2021, pp. 1--5.

\bibitem{sun2022q}
Heming Sun, Lu~Yu, and Jiro Katto,
\newblock ``Q-lic: Quantizing learned image compression with channel splitting,''
\newblock {\em IEEE Transactions on Circuits and Systems for Video Technology}, 2022.

\bibitem{shi2023rate}
Junqi Shi, Ming Lu, and Zhan Ma,
\newblock ``Rate-distortion optimized post-training quantization for learned image compression,''
\newblock {\em IEEE Transactions on Circuits and Systems for Video Technology}, 2023.

\bibitem{jeon2023integer}
Geun-Woo Jeon, SeungEun Yu, and Jong-Seok Lee,
\newblock ``Integer quantized learned image compression,''
\newblock in {\em 2023 IEEE International Conference on Image Processing (ICIP)}. IEEE, 2023, pp. 2755--2759.

\bibitem{cai2020rethinking}
Zhaowei Cai and Nuno Vasconcelos,
\newblock ``Rethinking differentiable search for mixed-precision neural networks,''
\newblock in {\em Proceedings of the IEEE/CVF Conference on Computer Vision and Pattern Recognition}, 2020, pp. 2349--2358.

\bibitem{sun2022entropy}
Zhenhong Sun, Ce~Ge, Junyan Wang, Ming Lin, Hesen Chen, Hao Li, and Xiuyu Sun,
\newblock ``Entropy-driven mixed-precision quantization for deep network design,''
\newblock {\em Advances in Neural Information Processing Systems}, vol. 35, pp. 21508--21520, 2022.

\bibitem{liu2022instance}
Zhenhua Liu, Yunhe Wang, Kai Han, Siwei Ma, and Wen Gao,
\newblock ``Instance-aware dynamic neural network quantization,''
\newblock in {\em Proceedings of the IEEE/CVF Conference on Computer Vision and Pattern Recognition}, 2022, pp. 12434--12443.

\bibitem{shi2023lossy}
Yumeng Shi, Shihao Bai, Xiuying Wei, Ruihao Gong, and Jianlei Yang,
\newblock ``Lossy and lossless (l2) post-training model size compression,''
\newblock in {\em Proceedings of the IEEE/CVF International Conference on Computer Vision}, 2023, pp. 17546--17556.

\bibitem{nagel2020up}
Markus Nagel, Rana~Ali Amjad, Mart Van~Baalen, Christos Louizos, and Tijmen Blankevoort,
\newblock ``Up or down? adaptive rounding for post-training quantization,''
\newblock in {\em International Conference on Machine Learning}. PMLR, 2020, pp. 7197--7206.

\bibitem{wei2022qdrop}
Xiuying Wei, Ruihao Gong, Yuhang Li, Xianglong Liu, and Fengwei Yu,
\newblock ``Qdrop: Randomly dropping quantization for extremely low-bit post-training quantization,''
\newblock {\em arXiv preprint arXiv:2203.05740}, 2022.

\bibitem{esser2019learned}
Steven~K Esser, Jeffrey~L McKinstry, Deepika Bablani, Rathinakumar Appuswamy, and Dharmendra~S Modha,
\newblock ``Learned step size quantization,''
\newblock {\em arXiv preprint arXiv:1902.08153}, 2019.

\bibitem{lee2021network}
Junghyup Lee, Dohyung Kim, and Bumsub Ham,
\newblock ``Network quantization with element-wise gradient scaling,''
\newblock in {\em Proceedings of the IEEE/CVF conference on computer vision and pattern recognition}, 2021, pp. 6448--6457.

\bibitem{jeon2022mr}
Yongkweon Jeon, Chungman Lee, Eulrang Cho, and Yeonju Ro,
\newblock ``Mr. biq: Post-training non-uniform quantization based on minimizing the reconstruction error,''
\newblock in {\em Proceedings of the IEEE/CVF Conference on Computer Vision and Pattern Recognition}, 2022, pp. 12329--12338.

\bibitem{begaint2020compressai}
Jean B{\'e}gaint, Fabien Racap{\'e}, Simon Feltman, and Akshay Pushparaja,
\newblock ``Compressai: a pytorch library and evaluation platform for end-to-end compression research,''
\newblock {\em arXiv preprint arXiv:2011.03029}, 2020.

\bibitem{lin2014microsoft}
Tsung-Yi Lin, Michael Maire, Serge Belongie, James Hays, Pietro Perona, Deva Ramanan, Piotr Doll{\'a}r, and C~Lawrence Zitnick,
\newblock ``Microsoft coco: Common objects in context,''
\newblock in {\em Computer Vision--ECCV 2014: 13th European Conference, Zurich, Switzerland, September 6-12, 2014, Proceedings, Part V 13}. Springer, 2014, pp. 740--755.

\bibitem{kodak1993kodak}
Eastman Kodak,
\newblock ``Kodak lossless true color image suite (photocd pcd0992),''
\newblock {\em URL http://r0k. us/graphics/kodak}, vol. 6, 1993.

\bibitem{asuni2013testimages}
Nicola Asuni and Andrea Giachetti,
\newblock ``Testimages: A large data archive for display and algorithm testing,''
\newblock {\em Journal of Graphics Tools}, vol. 17, no. 4, pp. 113--125, 2013.

\bibitem{CLIC2020}
George Toderici, Wenzhe Shi, Radu Timofte, Johannes~Balle Lucas~Theis, Eirikur Agustsson, Nick Johnston, and Fabian Mentzer,
\newblock ``Workshop and challenge on learned image compression (clic2020),'' 2020.

\end{thebibliography}

\newpage

\newpage
\appendix

\subsection{Coding performance of quantized LIC models}
\label{sec:appendix_1}

We compare the performance of our FMPQ method vs 8-bit FPQ on three LIC models: \textit{Scale Hyperprior} \cite{balle2018variational}, \textit{Mean Scale Hyperprior} \cite{minnen2018joint} and \textit{Cheng Anchor 2020} \cite{cheng2020learned}. We obtain both the fixed-precision quantized and mixed-precision quantized models from the pre-trained floating point baselines. After quantization, we fine-tune both the quantized models using the COCO dataset for 30 epochs with the Adam optimizer and use a learning rate of $10^{-5}$ for the model parameters and learning rate of $10^{-4}$ for the quantization parameters. We evaluate the performance of each of the three LIC models using three image datasets; Kodak \cite{kodak1993kodak}, Tecnick \cite{asuni2013testimages} and Clic \cite{CLIC2020}. In the following sections we show the Rate-Distortion (PSNR vs bpp) plot for each of the three LIC models using each of the three datasets. 

% \section{Scale Hyperprior Model}

\begin{figure*}
     \centering
     \begin{subfigure}[b]{0.326\textwidth}
         \centering
         \includegraphics[width=0.93\textwidth]{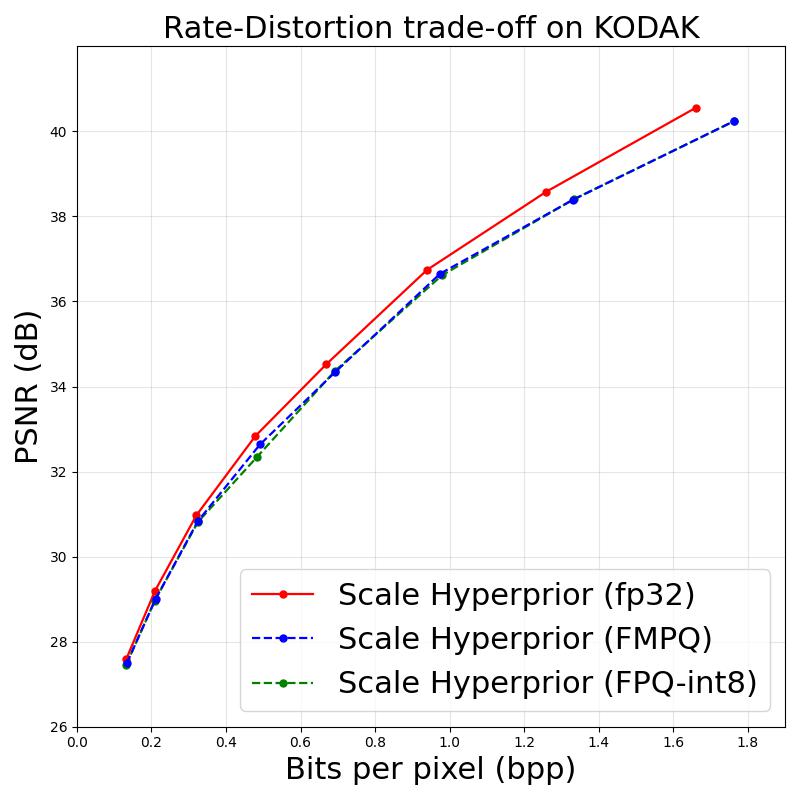}
         \caption{Scale Hyperprior, KODAK}
         \label{fig:sh-kodak}
     \end{subfigure}
     % \hfill
     \begin{subfigure}[b]{0.326\textwidth}
         \centering
         \includegraphics[width=0.93\textwidth]{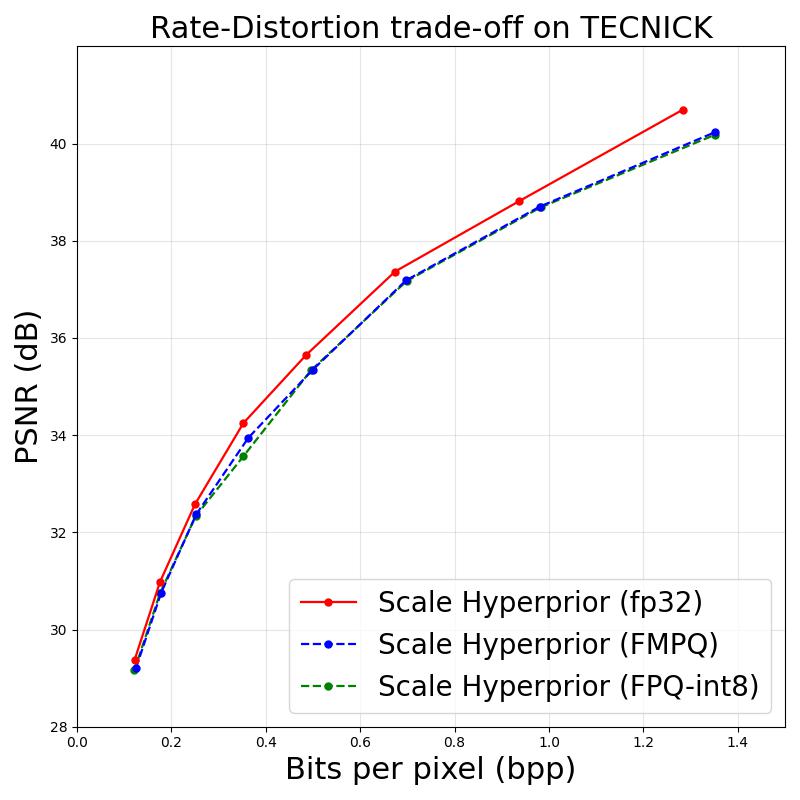}
         \caption{Scale Hyperprior, TECNICK}
         \label{fig:sh-tecnick}
     \end{subfigure}
     \begin{subfigure}[b]{0.326\textwidth}
         \centering
         \includegraphics[width=0.93\textwidth]{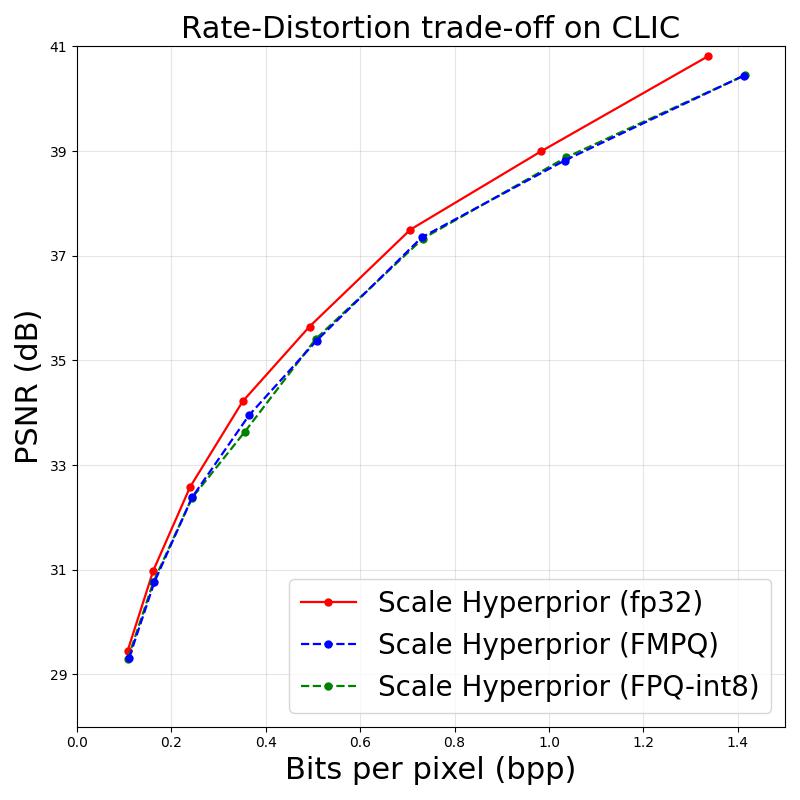}
         \caption{Scale Hyperprior, CLIC}
         \label{fig:sh-clic}
     \end{subfigure}
\vspace{-0.2cm}
\caption{Rate-Distortion (PSNR vs bpp) curves for the full-precision (\textit{float32}), FPQ (\textit{int8}) and proposed FMPQ models using the \textit{Scale Hyperprior} LIC model obtained using the KODAK, TECNICK and CLIC image datasets.}
\label{fig:RD-SH}
\vspace{-0.2cm}
\end{figure*}

In Fig.~\ref{fig:RD-SH}, we have the Rate-Distortion (PSNR vs bpp) curves for the \textit{Scale Hyperprior} LIC model. Each plot contains the curve for the full-precision (\textit{float32}), FPQ (\textit{int8}) and proposed FMPQ network. Fig.~\ref{fig:sh-kodak} contains the curves obtained using the KODAK dataset, Fig.~\ref{fig:sh-tecnick} contains the curves for the TECNICK dataset and Fig.~\ref{fig:sh-clic} contains the curves for the CLIC dataset. These Rate-Distortion curves were used to obtain the BD-Rate values of Table.~\ref{tab:tab2}.

% \subsection{Mean Scale Hyperprior Model}

\begin{figure*}
     \centering
     \begin{subfigure}[b]{0.326\textwidth}
         \centering
         \includegraphics[width=0.93\textwidth]{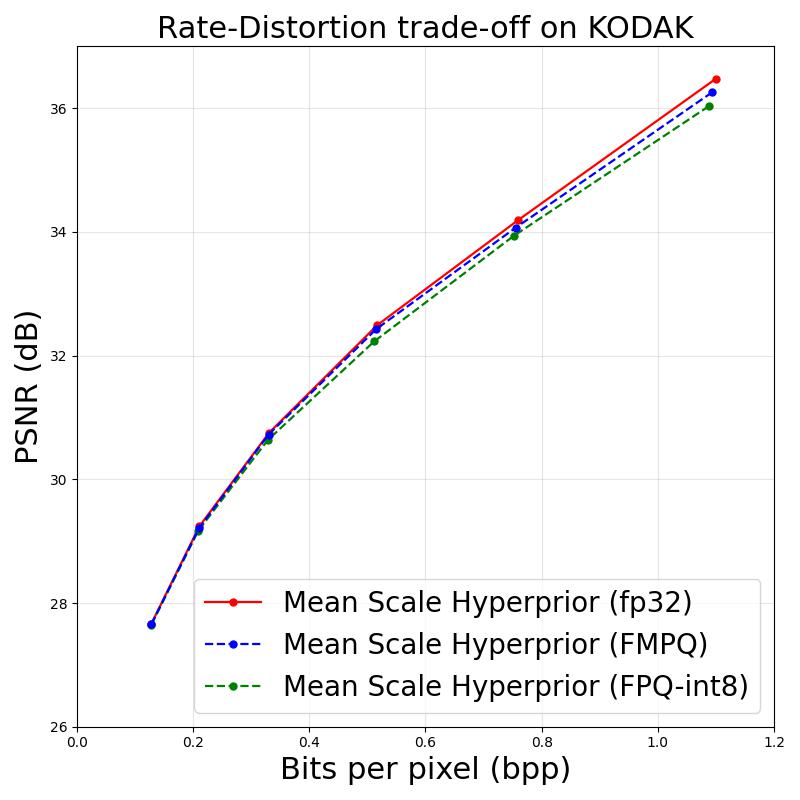}
         \caption{Mean Scale Hyperprior, KODAK}
         \label{fig:msh-kodak}
     \end{subfigure}
     % \hfill
     \begin{subfigure}[b]{0.326\textwidth}
         \centering
         \includegraphics[width=0.93\textwidth]{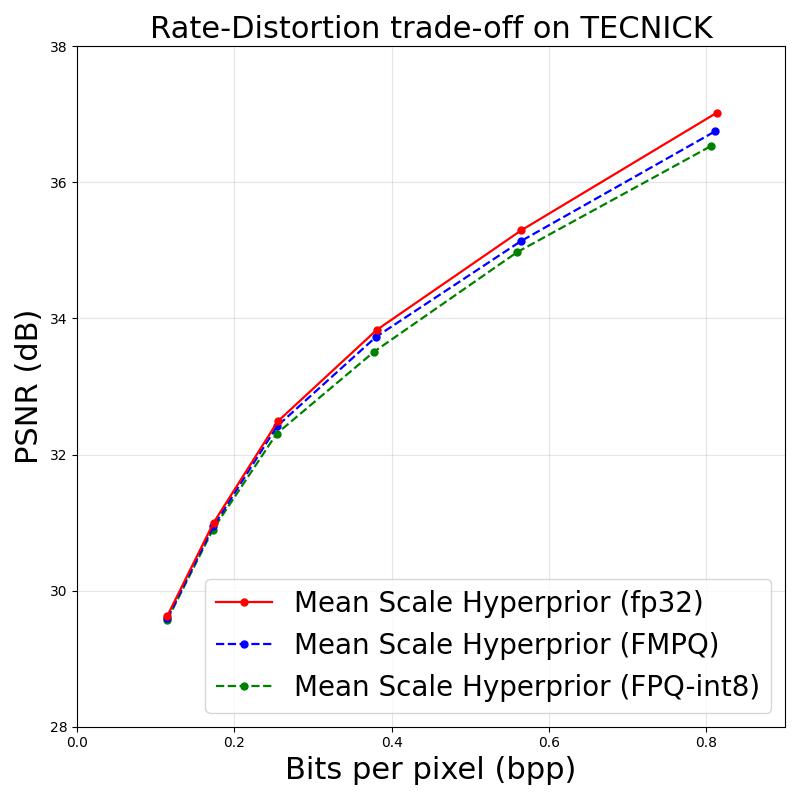}
         \caption{Mean Scale Hyperprior, TECNICK}
         \label{fig:msh-tecnick}
     \end{subfigure}
     \begin{subfigure}[b]{0.326\textwidth}
         \centering
         \includegraphics[width=0.93\textwidth]{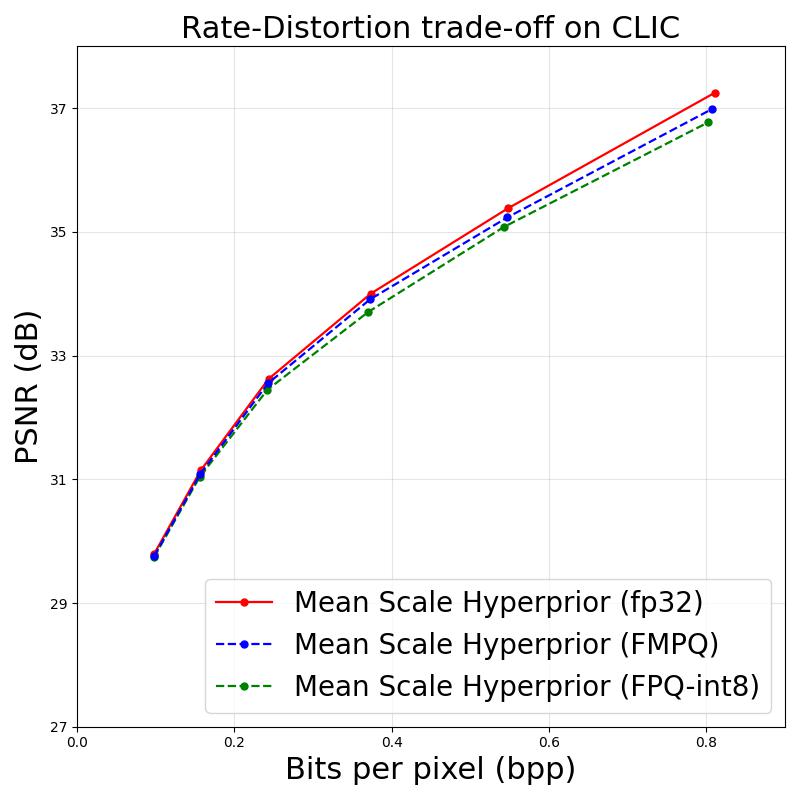}
         \caption{Mean Scale Hyperprior, CLIC}
         \label{fig:msh-clic}
     \end{subfigure}
\vspace{-0.2cm}
\caption{Rate-Distortion (PSNR vs bpp) curves for the full-precision (\textit{float32}), FPQ (\textit{int8}) and proposed FMPQ models using the \textit{Mean Scale Hyperprior} LIC model obtained using the KODAK, TECNICK and CLIC image datasets.}
\label{fig:RD-MSH}
\vspace{-0.2cm}
\end{figure*}

In Fig.~\ref{fig:RD-MSH}, we have the Rate-Distortion (PSNR vs bpp) curves for the \textit{Mean Scale Hyperprior} LIC model. Each plot contains the curve for the full-precision (\textit{float32}), FPQ (\textit{int8}) and proposed FMPQ network. Fig.~\ref{fig:msh-kodak} contains the curves obtained using the KODAK dataset, Fig.~\ref{fig:msh-tecnick} contains the curves for the TECNICK dataset and Fig.~\ref{fig:msh-clic} contains the curves for the CLIC dataset. These Rate-Distortion curves were used to obtain the BD-Rate values of Table.~\ref{tab:tab2}.

% \subsection{Cheng Anchor Model}

\begin{figure*}
     \centering
     \begin{subfigure}[b]{0.326\textwidth}
         \centering
         \includegraphics[width=0.93\textwidth]{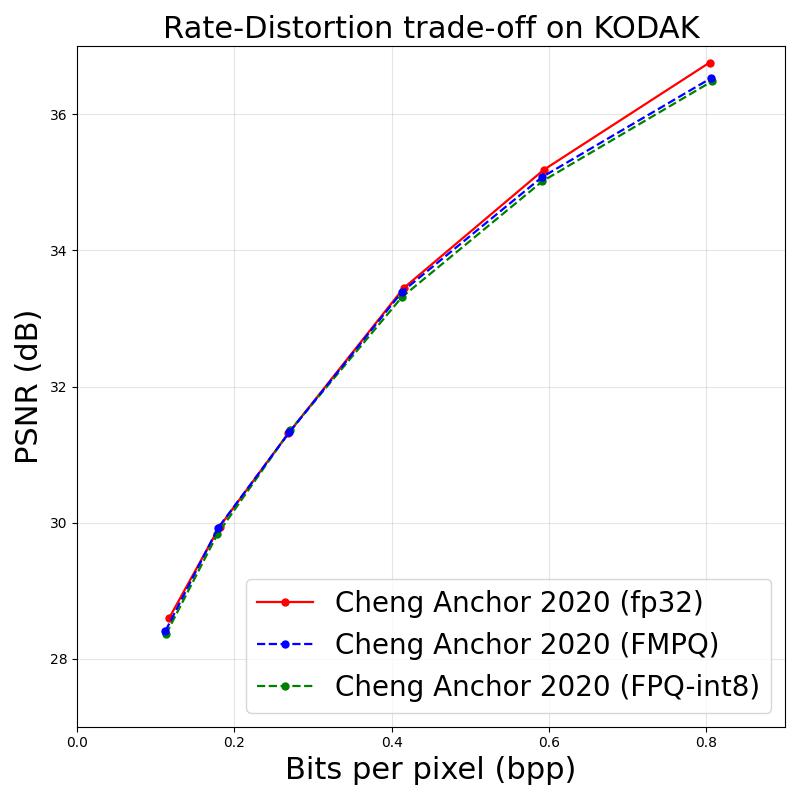}
         \caption{Cheng Anchor, KODAK}
         \label{fig:ca-kodak}
     \end{subfigure}
     % \hfill
     \begin{subfigure}[b]{0.326\textwidth}
         \centering
         \includegraphics[width=0.93\textwidth]{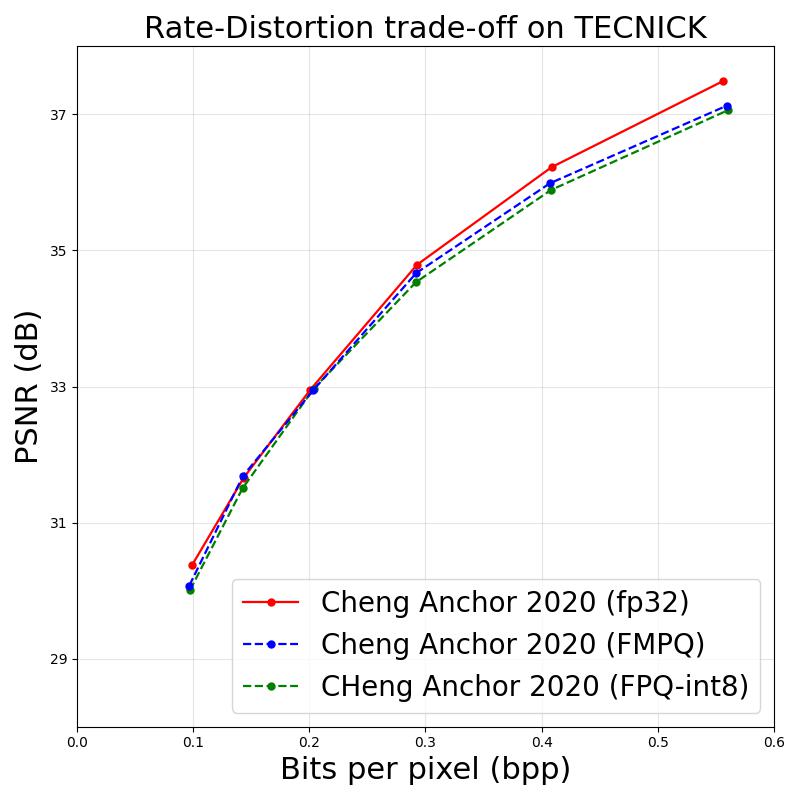}
         \caption{Cheng Anchor, TECNICK}
         \label{fig:ca-tecnick}
     \end{subfigure}
     \begin{subfigure}[b]{0.326\textwidth}
         \centering
         \includegraphics[width=0.93\textwidth]{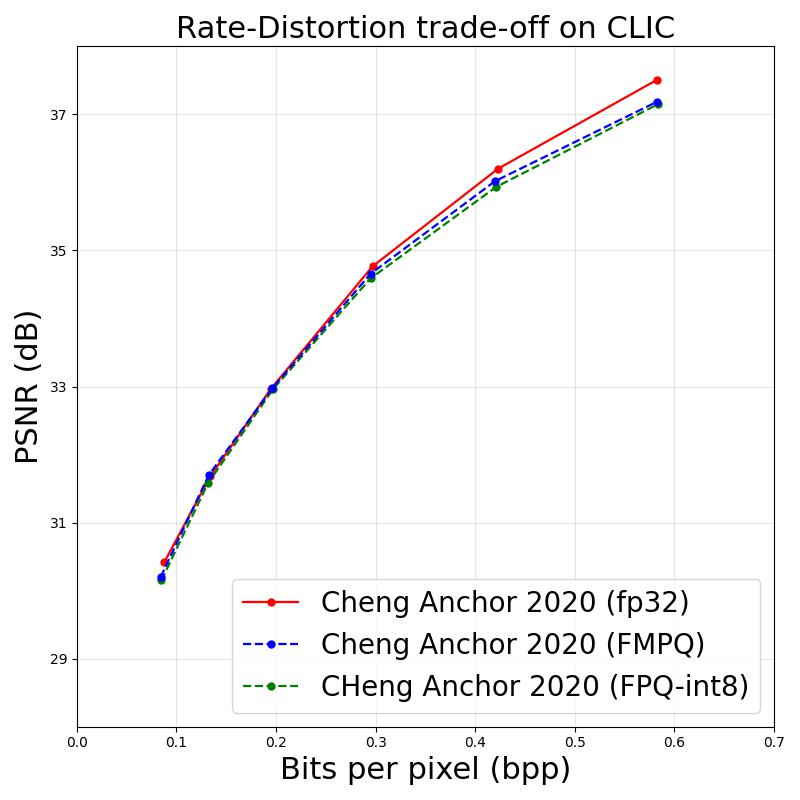}
         \caption{Cheng Anchor, CLIC}
         \label{fig:ca-clic}
     \end{subfigure}
\vspace{-0.2cm}
\caption{Rate-Distortion (PSNR vs bpp) curves for the full-precision (\textit{float32}), FPQ (\textit{int8}) and proposed FMPQ models using the \textit{Cheng Anchor 2020} LIC model obtained using the KODAK, TECNICK and CLIC image datasets.}
\label{fig:RD-CA}
\end{figure*}

In Fig.~\ref{fig:RD-CA}, we have the Rate-Distortion (PSNR vs bpp) curves for the \textit{Cheng Anchor} LIC model. Each plot contains the curve for the full-precision (\textit{float32}), FPQ (\textit{int8}) and proposed FMPQ network. Fig.~\ref{fig:ca-kodak} contains the curves obtained using the KODAK dataset, Fig.~\ref{fig:ca-tecnick} contains the curves for the TECNICK dataset and Fig.~\ref{fig:ca-clic} contains the curves for the CLIC dataset. These Rate-Distortion curves were used to obtain the BD-Rate values of Table.~\ref{tab:tab2}.

\subsection{Model Size calculations}
\label{sec:appendix_3}

The model sizes reported in the paper are the averages across all the networks of the LIC models corresponding to different quality levels. While calculating the model size for a quantized layer, we consider the quantization parameters $s$ and $w$ are stored in \textit{float32} precision. We use Eq.~\ref{eqn:4} to calculate the size in \textbf{MB}, $M$ of a convolution layer with shape $(C_{out}, C_{in}, k, k)$.
\vspace{-0.4cm}

\begin{equation}  \label{eqn:4}
M = (C_{out} \times C_{in} \times k^2 + C_{out}) \times b + C_{out} \times 2 \times 32
\end{equation}
Here $b$ is the assigned bit-width for the layer. The first term represents the memory requirements for the model parameters and the second term represents the memory requirements for the quantization parameters.

\subsection{BD-Rate vs Model Size Trade-off obtained using FMPQ Method}
\label{sec:appendix_2}

\begin{figure}[H]
 \centerline{\epsfig{figure=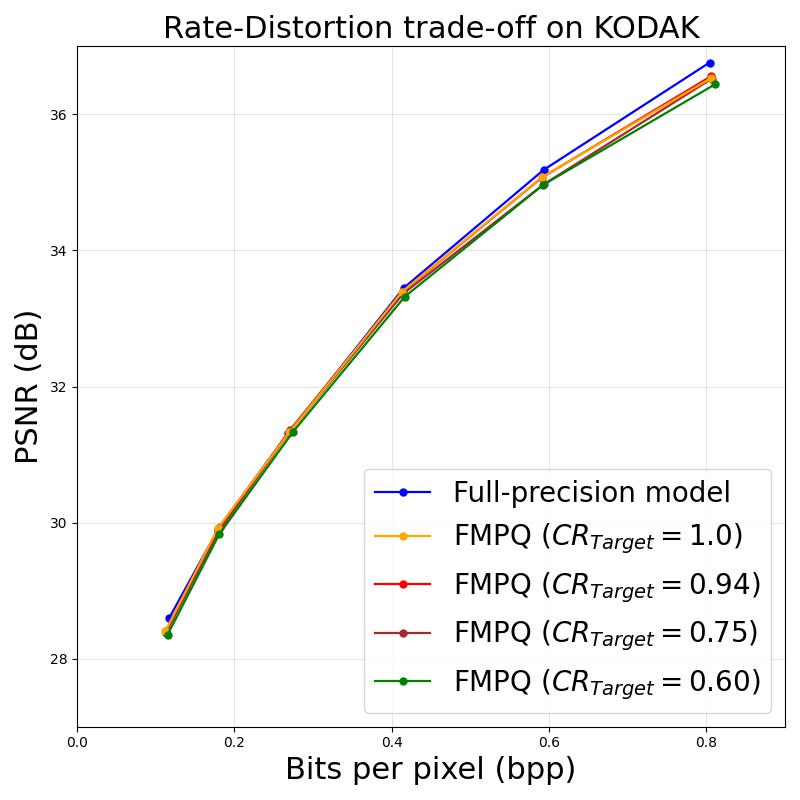, width=1.0\linewidth}}
\caption{Rate-Distortion curves obtained for different values of the hyperparameter $CR_{\textit{Target}}$ using the \textit{Cheng Acnhor} LIC model and KODAK dataset.}
\vspace{-0.4cm}
\label{fig:rd_flexible}
\end{figure}

Fig.~\ref{fig:rd_flexible} contains Rate-Distortion curves used to ontain the results of Table.~\ref{tab:tab3}. The figure contains five Rate-Distortion curves and is obtained by using the \textit{Cheng Acnhor} LIC model evaluated on the KODAK dataset. The blue line corresponds to the full-precision network while the other lines correspond to FMPQ networks obtained using different values of the hyperparameter $CR_{\textit{target}}$. We can see that as $CR_{\textit{target}}$ is reduced (to obtain a quantized model with smaller size), the Rate-Distortion curve moves further away from the full-precision network curve resulting into a lower BD-Rate. By varying $CR_{\textit{target}}$, we can obtain a family of Rate-Distortion curves. In practical settings, we can choose the curve that fits our model size constraints and operate on a point in that curve that satisfies our bit-rate limitations or image quality requirements.

\end{document}